# Authorization Policies and Co-Operating Strategies of DSCloud Platform

Lican Huang
Hangzhou Domain Zones Technology Co., Ltd
Hangzhou, China
huang_lican@yahoo.co.uk, licanhuang@zstu.edu.cn

*Abstract* — One of the services of DSCloud Platform is to provide the global directory service to solve the problems of dispersed, difficult retrieved and isolated information. In this paper, we describe DSCloud Platform's authorization policies and co-operating strategies for articles and comments, and usage scenery for co-editing posts and tables in the platform.

## I. INTRODUCTION

In the Clouds or Web sites, people publish the posts (upload files). There are requirements to protect the rights such as accessing, publishing, and auditing the posts (files). Today's authorization policies are not enough to control the users' all activities.

DSCloud platform[1] is a software whose final goal is to provide various services for global users based on semantic P2P networks[2,3,4,5,6,7,8], in which all information is classified into hierarchical domains. DSCloud platform provides a global directory service by which the directories are created and maintained by the users themselves. For DSCloud platform, users can create directories, publish posts(upload files), and comments. We need to control in detail the users' read, write, audit activities and co-operating such co-editing Excel-like table in the Web.

This paper presents authorization policies and co-operating strategies of DSCloud Platform.

## II. THE PHILOSOPHY OF AUTHORIZATION DESIGN OF DSCLOUD PLATFORM

DSCloud Platform aims to build a platform to provide various services for millions users. We hope it is managed automatically by users themselves, and by global control as least as possible. As provided services such as communication services, education services, office suite services, we hope the users can create and manage their own services, for example, online test question database.

DSCloud Platform provides directory service, which the user can create sub directories. In the meantime, we hope the directory has choices of prohibit or permit to do some actives such as creating new directories, reading papers in the directories and so on. We give these choice controls to the owner of the directory(the creator of the directory). In the same way, for the papers (posts, files), we give the choice controls to the owner of the paper(post, file). The global control only happens when the users do unlawful actions in the platform. In these cases, we can block the users' activities and their created directories or papers.

## III. AUTHORIZATION POLICIES OF DSCLOUD PLATFORM

Authorization polices of DSCloud Platform are matrix table in which first column are as user's roles and first row as user's authorizations. The user's roles include creator of directory(domain), creator of file (post), anyuser, thisGroup-within the directory(domain) group, grantGroup-within the granted group, and granted user. The group has two types, public and private. Any user can join public group, but for private group, the user requests for joining must be reproved by the owner of the group. In DSCloud Platform, the group names, domains or directories are same.

### 3.1 Authorization Polices of Directory

In DSCloud Platform, the user who creates the directory is the creator and owner of the directory. The owner can delete the directory when there are no joined users and no published articles in this directory. The owner can also trash the directory. The owner can set the authorization of the directory created by himself.

The five roles are DirCreator, thisGroup, grantGroup, grantUser, and AnyUser; and four rights are publish, Read, Create Sub Dir and Show Dir. The details of information are in the paper[9].

### 3.2 Authorization Polices of Files(Posts)

The user who creates the file is called as the owner or creator the file. The owner or creator of the file can setup the file authorization policies. Authorization polices of files are matrix as well. The five roles are the same as authorization polices of directory. However, the authorizations are Audit, Read, Write and Execute. "Audit" means the user with this right can audit the file(post), that is, can block or approve the file. "Read" means the user with this right can read (access) the file (post). "Write" means the user with this right can

edit(modify) the file(post)., and "Execute" is for the future use.

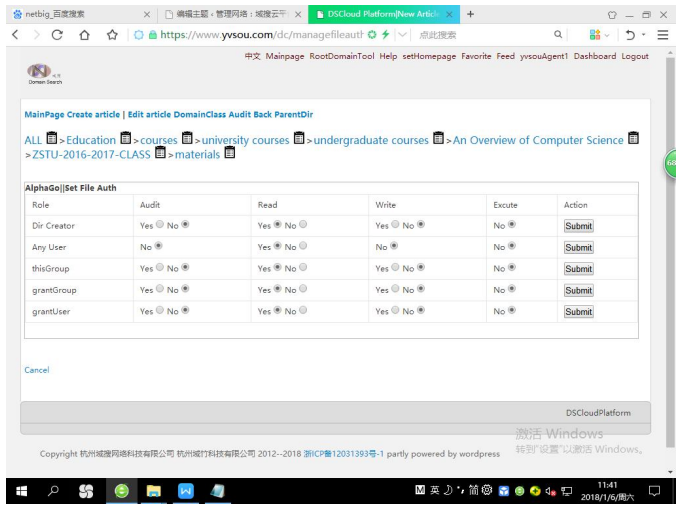

Fig.1 Setup Authorization Polices of Files

The user may have more than one role among 5 roles. If user's any role has the right, then the user has this right. Because the "read" right of directory controls all the posts (files) in the directory, the file can be accessed only when the rights of "read" for both directory and file are true.

### 3.3 Authorization Polices of Comments

There are different requirements for comments. Authorization polices of Comments of DSCloud Platform are matrix as well. The five roles and four authorizations are the same as authorization polices of file. However, the meanings of the authorizations are different. "Audit" means the user with this right can audit the comments, that is, can block the comments or approve the comments. "Read" means the user with this right can read (access) the comments. "Write" means the user with this right can publish the comments, and "Execute" is for the future use.

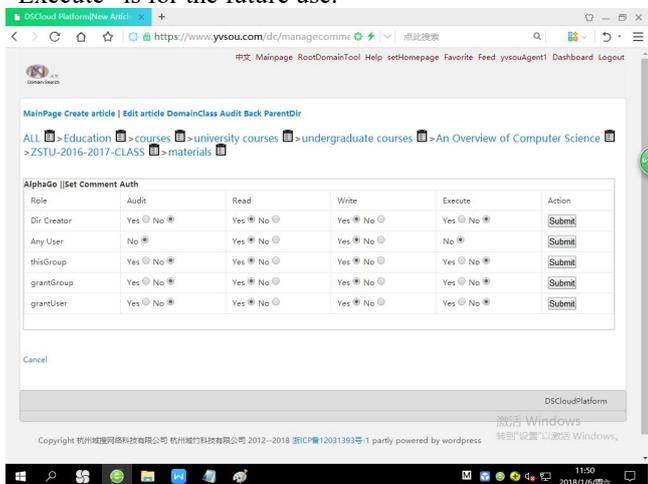

Fig.2 Setup Authorization Polices of Comments

## IV. CO-OPERATING STRATEGIES

Some times, the users in the group may co-edit the file (post). In DSCloud Platform, it is easy to implement this function. The directory owner setups the directory as private. The owner of the post in this directory setups the authorization policies to give thisGroup the "Write" right. Then, the users who want to co-edit the post applies to join the directory group. And the directory owner approves these users. Then all users in this group can co-edit the post. In the case when the owner of the directory and the owner of the post in the directory is the same, it is very easy to setup all these steps. Another usage for co-editing in the Web is co-editing Excel-like table in the Web.

### 4.1 Excel Co-editong by Using Handsontable JS

The creator of the post create a new post with shortcode as the following:
[excelcode]this is excel co editing example1 [/excelcode]

When the post published, the owner read the post, the following figure 3 will be shown. Then importing the local excel file into the web.

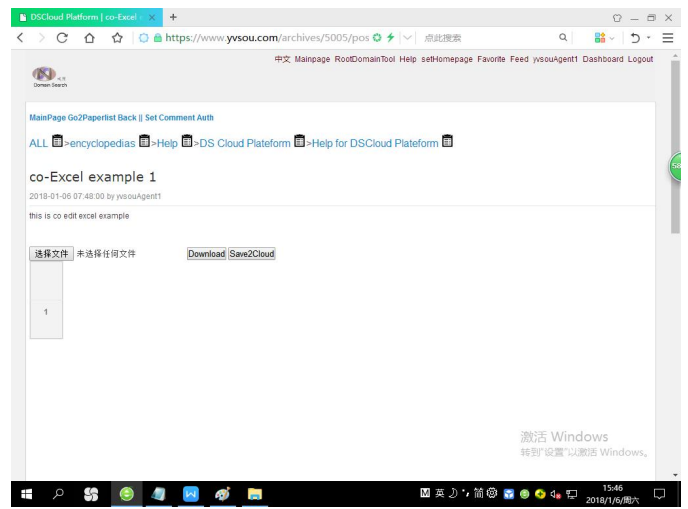

Fig3 Co-edit Excel file in the Web before importing file

When importing the local file, and saving file to DSCloud (save2cloud), then in the next time, the web will be shown as figure 4.

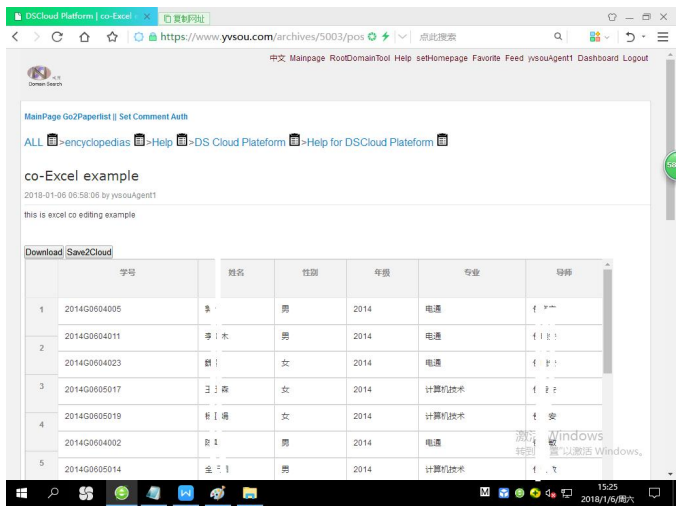

Fig4. Co-edit Excel file in the Web after save file to DSCloud

The user with 'Write' right can modify the excel table, and can download the excel table into local file. If the user only with "Read" right, the excel table can not be modified, the user can only see the Excel table.

We implemented the co-edit excel table in the web by using JS-XLSX [10] and Handsontable JS [11].

4.2  More Level Right Control for Co-editing Table

In 4.1, the authorization policies of tables are taking the table as a whole. This is, the table as a whole can be or not be accessed by the users. However, there are many cases in which the rows or columns are required to be controlled by the authorization policies. We designed another short-code to satisfy these needs.
The shortcode of Wordpress[12] is like the following:

[cotablecode tid =2 colshow ="011111" headcols="sid, name, sex, email, telephone, hobbies "  rights="F7775" ] This is an example for controlling all rows and cols  [/cotablecode]

Here, tid is for the table no. in the post; colshow means that the column will be hidden if the character with related position is zero. For example, colshow ="011111" means that the first column will be hidden. Headcols is the table columns; rights is for the authorization policy matrix. The default values are as "tid" => '1', "headcols"=> ","colshow" => '1111111111', "rights" => '77775' . When headcol is empty, then a local Excel file will be required to be imported.

The "rights" is for 5 roles (fileOwner, thisGroup, grantGroup, grantUser, and AnyUser). rights="F7775" means that fileOwner has "F" right, thisGroup, grantGroup, grantUser with "7" right, and anyUser with "5" right. The 4 rights are "whole read", "read", "write", and "add", which is expressed like Linux file rights. That is, the rights are Hexadecimal String, in which each character is 4 bits like 1010. Rows may be added and updated by different users. There are requirements to control access of rows and columns.

"whole read" means the user with this right can see all rows and columns of the tables. "read" means the user with this right can see all rows and columns of the tables, but some columns may be hidden by colshow. "write" means the user with this right can update and delete the rows, and "add" means the user with this right can add a new row. In the meanwhile, the owner of the row ( the user who adds the row) has all 4 rights, which means the owner of the row can do anythings for his own rows, he can read , update and delete his own rows.

When the above shortcode is published within a post, then the post will be shown as the Fig.5. Then other users with the related rights can add , update and delete the rows of the tables.

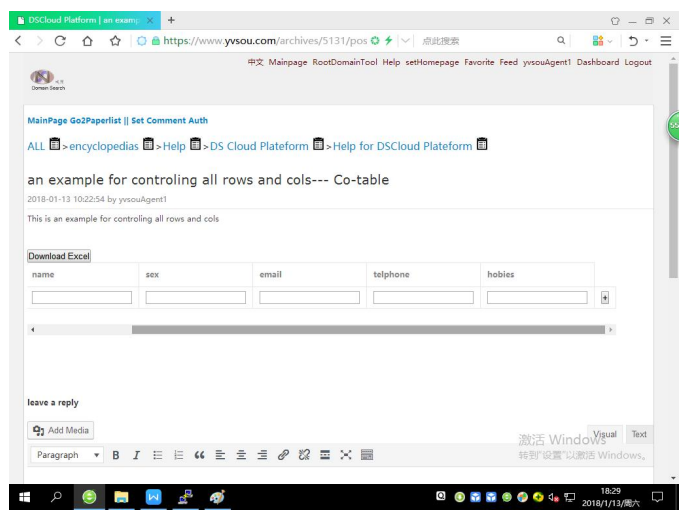

Fig5. Co-editing table in the Web for rows and columns

V.  CONCLUSIONS

This paper presents DSCloud Platform's philosophy about autonomous managements by users themselves, and authorization policies and co-operating strategies for directories, articles, comments, and tables. These authorization policies may be used for large distributed systems.

ACKNOWLEDGMENT

The paper is supported by the project "Hangzhou Qinglan Plan--Scientific and technological creation and development (No.201318 31K99" of Hangzhou scientific and technological committee. The software copyrights is owned by Hangzhou Domain Zones Technology Co., Ltd. By agreements, Chinese patent applied is owned by Hangzhou Domain Zones Technology Company.